# Harnessing behavioral diversity to understand circuits for cognition

Simon Musall[1], Anne Urai[1], David Sussillo[2,3,4], Anne Churchland[1]

1. Cold Spring Harbor Laboratory, Cold Spring Harbor, NY
2. Google AI, Google Inc., Mountain View, CA, USA
3. Department of Electrical Engineering, Stanford University, Stanford, CA, USA.
4. Stanford Neurosciences Institute, Stanford University, Stanford, CA, USA

# Harnessing behavioral diversity to understand circuits for cognition

## Abstract

With the increasing acquisition of large-scale neural recordings comes the challenge of inferring the computations they perform and understanding how these give rise to behavior. Here, we review emerging conceptual and technological advances that begin to address this challenge, garnering insights from both biological and artificial neural networks. We argue that neural data should be recorded during rich behavioral tasks, to model cognitive processes and estimate latent behavioral variables. Careful quantification of animal movements can also provide a more complete picture of how movements shape neural dynamics and reflect changes in brain state, such as arousal or stress. Artificial neural networks (ANNs) could serve as an important tool to connect neural dynamics and rich behavioral data. ANNs have already begun to reveal how particular behaviors can be optimally solved, generating hypotheses about how observed neural activity might drive behavior and explaining diversity in behavioral strategies.

# Introduction

To understand the computations implemented by neural circuits, it is critical to study them in the context of the behavioral output they generate. Studies of decision-making usually leverage behavioral tasks in which animals are trained to produce a specific behavioral response following presentation of sensory stimuli. This allows repeated measurements of neural activity with systematic manipulation of the inputs, supporting the ability to map the neural pathways that transform sensory inputs into action [1–4]. However, focusing only on binary choices might cause one to overlook additional ongoing behaviors as well as the animal's brain state, both of which strongly affect neural activity and task performance [5–10]. Simple tasks, such as licking in response to a sensory stimulus, may also constrain the dimensionality of the observed neural dynamics (Box 1), making it difficult to estimate whether they accurately represent neural function under more complex conditions [11–13].

Recent experimental, analytical and theoretical advances provide opportunities to overcome these issues, bolstering our ability to connect neural activity to function. Here, we provide an overview of emerging methods and argue that the decision-making field should embrace behavioral complexity as a way to understand apparently spontaneous fluctuations in neural activity, gain insight into an animal's brain state and distinguish behavioral strategies. First, we describe task features that increase behavioral complexity and allow one to infer an animal's estimate of computationally relevant quantities (e.g., weight of evidence in favor of a specific choice). Such latent behavioral variables are not directly measured but can be derived from behavioral models. Second, we highlight new ways of quantifying animal movements and behavioral motifs and describe how such data can aid the interpretation of observed behavior and task strategy (Fig. 1), as well as single-trial neural data. Lastly, we highlight the use of artificial neural neural networks (ANNs), especially multi-layered networks also known as deep neural networks, as a way to create hypotheses for how high-dimensional neural dynamics can give rise to behavior. ANNs can be viewed as simple artificial model organisms for which the entire connectome and activation space is known. Knowledge of behavioral latent variables can be linked to internal ANN dynamics underlying a range of task-relevant computations [4,14–19]. By including more detailed behavioral quantification and task complexity, ANN outputs and task strategy could be further constrained to generate network dynamics that are more comparable to biological neural circuits. ANNs are also starting to provide insight into individual differences in behavioral strategy, even for animals that pursue the same experimental goal (Fig. 1) [20].

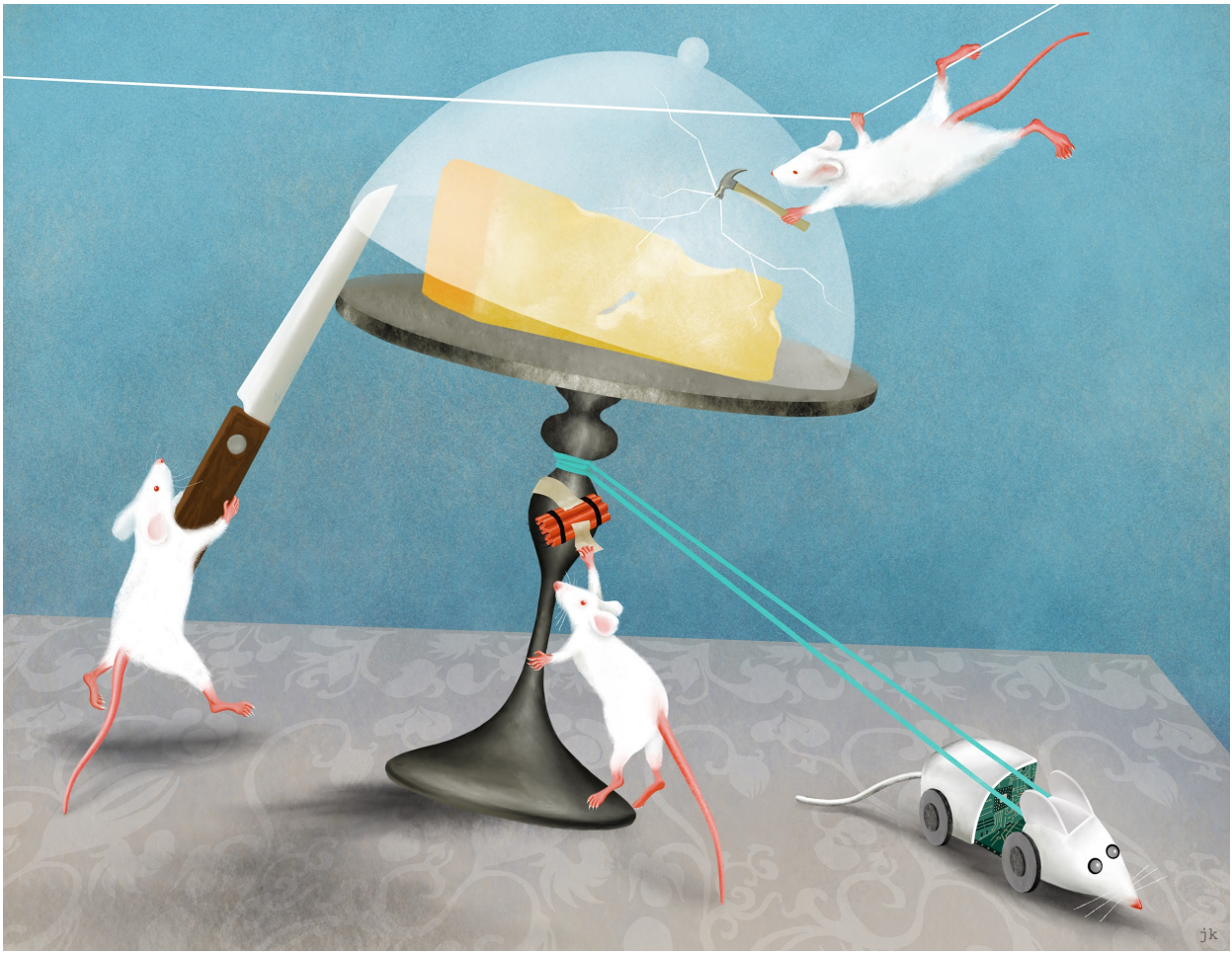

*Figure 1. Animals can exhibit a diverse range of behaviors and strategies even when solving the same task. Insight into this diversity might come from increasing task complexity, detailed quantification of animal behavior and examination of ANNs trained to solve the same problem.*

## Task features to estimate latent behavioral variables

In the study of decision-making, animals often perform a behavioral task designed to engage a cognitive process of interest. This *task-based approach* can support the extraction of *latent behavioral variables* (Table 1), which are not directly observable from behavior but inferred through mathematical models of cognitive processes. Latent variables explain variability in animal choices [21] which can be related to neural activity. But which features should be incorporated in a behavioral task to effectively estimate latent variables? Three task domains can be informative about latent behavioral variables: Stimulus presentation, animal responses and task structure.

Multiple stimulus features can be used by experimenters to fit models that estimate latent behavioral variables. For instance, stochastic, time-varying stimuli allow the use of behavioral models to infer the animal's time-varying estimate of accumulated evidence; these can then be related to neural activity [22,23]. Presenting different sensory modalities at varying levels of reliability allows one to estimate their respective weight during multisensory integration [24] and evaluate how these weights are encoded in population neural activity [25].

An animal's responses can likewise be used to infer latent behavioral variables. For instance, reaction times provide an estimate of the time to reach a decision bound in models of evidence accumulation [26,27] and post-choice waiting times are linked to decision confidence [28–30]. Continuous responses like reaching movements, moving a wheel or navigating in virtual reality can provide additional insight into the evolving decision process [31–34]. For example, changes in head orientation are related to upcoming animal choices and trial-to-trial neural variability [35]. Further, reaching trajectories in a virtual maze task can reveal changes of mind in free-choice trials [31]. Online video tracking offers great promise to further expand the repertoire of continuous task-relevant behaviors, without the need for invasive procedures or custom hardware [36,37; see below].

In between stimulus and response lies a vast space of task design choices that can be used to study additional latent variables. Cognitive models can be fit to behavior and detail the algorithms by which agents may perform probabilistic inference [38], learn the structure of their environment [39] or adjust their decision policy after errors [40]. The model's latent variables can then be linked to activity in specific brain regions [41] and single cells [42]. Dynamic logistic regression models also allow data-driven estimation of across-trial latent variables, such as an animal's reliance on trial history or its choice bias, over long timescales and without assuming a specific generative model [43].

Box (separate) -----------------------

How can richer behavioral tasks, combined with increasingly high-dimensional neural measurements, help us to relate complex neural dynamics to behavior? The theory of neural task complexity (NTC) states that the dimensionality of neural population dynamics has an upper bound defined by the number of task parameters and the smoothness of neural trajectories across those parameters [44]. In simple tasks, neural network dynamics are therefore constrained by the low number of task parameters and contain far fewer dimensions compared to the number of recorded neurons. To overcome this issue, NTC can be used to compute the expected dimensionality of neural network-dynamics when increasing task complexity. This framework thus promises to be a valuable tool to titrate the complexity required for future behavioral tasks to match our growing capacity for recording many neurons simultaneously [12].

----------------------------

*Table 1: Definitions of new terms emerging in the study of cognitive and behavioral circuits*

| Term | Definition | Examples | Biological vs artificial networks? | |
|---|---|---|---|---|
| Brain state | Internally generated neural dynamics that fluctuate spontaneously, often in ways that are related to bodily constraints. These are often measured through physiological markers. | Arousal, fear, stress, hunger, motivation, engagement, drowsiness | Biological | 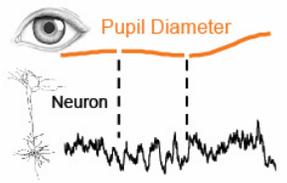 |
| Latent behavioral variable | The agent's estimate of a computationally relevant quantity. These are inferred via behavioral models. | Accumulated evidence, bias, value, confidence | Both | 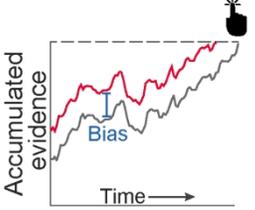 |
| Behavioral motif | A stereotyped series of movements that identifies a specific behavior. Ongoing behavior can be described as a continuous sequence, switching from one motif to the next. | Grooming, eating, mating, walking, reaching, rearing | Both | 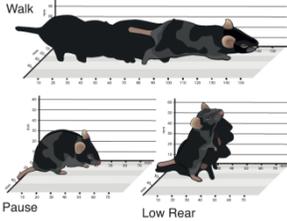 |
| Task-based approach | Experiments with an experimenter-defined task and a repeatable trial-structure. Non-human experiments usually include animals trained to perform an arbitrary movement to receive a reward. | Random dot motion task, reaching task, maze navigation task, image classification task | Both | 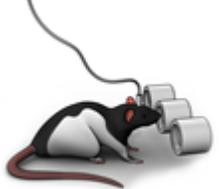 |
| Observational approach | Experiments where self-generated behavior of untrained animals is observed and analyzed. | Home cage exploration, mating behavior, head-fixed wheel locomotion, place field mapping | Biological [but see cite for ANN examples] | 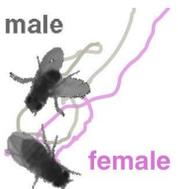 |
| Individual differences | Differences in an individual's behavioral repertoire that allow to distinguish different animal types. | Exploratory vs. fearful, dominant vs. submissive, social vs. asocial, active vs. inactive | Both | 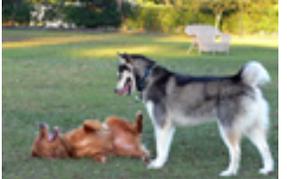 |

# Quantification of animal behavior

The earliest descriptions of animal behavior were generated by ethologists taking an *observational approach* (Table 1): they formulated a set of criteria (usually describing a sequence of simpler movements) to identify a specific *behavioral motif* (e.g. feeding or grooming; Table 1) and quantify its occurrence [45]. With the rise of neural recordings in freely-moving animals, such behavioral motifs are now routinely related to neural activity patterns [46,47]. While quantifying behavioral motifs used to be extremely laborious and susceptible to human error, technical advances have largely shifted the field towards automatic identification and quantification of pre-defined behavioral motifs [48].

There is a large toolbox to measure behavior, such as audio recording [49] or RFID tagging [50], but the most striking (and potentially most applicable for decision-making) methods are based on video data. New toolboxes utilize supervised (based on human-labeled examples) deep learning methods, to automatically track the position and posture of animals as they navigate through their environment [36,37]. Given enough training data, these algorithms are remarkably robust, and can be readily applied to video data to obtain readouts like movement velocity, spatial position, and body orientation.

An extension of animal movement quantification is the unsupervised classification of behavioral motifs without a preconceived, observation-based template. Different analyses have been optimized for species like C. elegans [51], larval zebrafish [52], and drosophila melanogaster [53], but also freely moving mice [54]. Here, time-frequency analysis or auto-regressive models can discover and quantify the occurrence of stereotypic temporal dynamics in low-dimensional movement representations and cluster them into distinct behavioral motifs. This unsupervised, data-driven approach can reveal the full range of an animal's behavioral motifs, their frequency of occurrence and sequential order [54,55], which can be linked to simultaneous neural recordings. The approach has also begun to overturn long-held assumptions about the complexity of behavior. For instance, unsupervised classification methods applied to large databases of drosophila songs exposed additional song modes beyond the two that were long thought to make up the animal's song repertoire [56].

It is tempting to assume that detailed movement quantification is only necessary in the observational approach, and that animals in a task-based setting mostly reproduce a simple set of instructed movements (e.g. licking to obtain a reward). However, even head-fixed mice execute a large array of movements [9,10]. These uninstructed movements include whisking, locomotion and facial movements, are easily quantified by dimensionality reduction of video data and strongly affect neural activity [5–10]. During a decision-making task, video-based movement representations are also closely related to neural population dynamics, outweighing the importance of task-related variables like sensory stimuli or animal choice to predict neural activity in single trials [10]. Given this large impact of uninstructed movements on neural activity, their accurate quantification is therefore critically important when analyzing neural data during decision-making.

# Movements and brain state

Quantifying animal movements also gives experimenters a handle on tracking fluctuations in *brain states* (Table 1). Brain state changes account for large, widespread fluctuations in neural excitability, interneuronal correlations, oscillatory power of local-field potentials, stimulus-response amplitude and task performance [5,6,57–61]. Despite their pronounced impact on neural activity, the full range of distinct brain states is not well characterized (especially in awake animals); this remains an area of active research [62].

Automatic recognition of behavioral motifs can be used to continuously infer brain states like fear [63] or stress [64,65]. During decision-making, such behavioral quantification might aid in interpreting long-term changes in choice behavior and neural activity. For example, over the course of learning, an animal might be stressed from exposure to a novel environment but then habituate over time. Corresponding behavioral changes might be correlated with task performance and drive changes in neural activity that could be confounded with task learning. Continuous behavioral tracking would be a powerful way to address this concern.

A particularly valuable (involuntary) movement to capture brain state changes is pupil diameter [5,6,59,61,66,67]. Fluctuations in pupil diameter are linked to release of acetylcholine and noradrenaline, with fluctuations below 0.3 Hz being more closely related to cholinergic release and higher frequencies to adrenergic release [68]. The utility of pupil diameter data might therefore be increased by using spectral analyses to isolate different neuromodulatory components that affect cortical processing. Phasic pupil-linked arousal can also reflect and interact with latent behavioral variables: for example, pupil dilation scales with decision confidence and reduces across-trial serial choice biases [69]. Future behavioral models should therefore use brain state measures (e.g. pupil dilations) to continuously adjust latent behavioral variables and improve their accuracy when predicting changes in neural activity and task performance.

Animal locomotion also has profound effects on neural activity. While sometimes used as an alternate measure of arousal [7,70,71], locomotion is mainly associated with a high arousal state and does not accurately reflect the full spectrum of state modulations seen with pupil measures [62]. Pupil dilations are also seen during quiescence and can be observed seconds before and after bouts of locomotion [6,68]. Accordingly, locomotion is usually followed by severe de- or hyperpolarization across cortical areas for prolonged periods of time [71]. Many sensory neurons are also modulated by locomotion alone, independent of brain state. Auditory neurons are inhibited by locomotion, likely to suppress the perception of predictable, movement-related sounds [72] and while firing rates of V1 neurons are suppressed during arousal, they are elevated during locomotion [6]. Combining at least 16 distinct movement dimensions (during head-fixation), also predicts far more variance in neural activity than a model based on state-related movements like pupil dilation, locomotion, and whisking alone [9]. This emphasizes the importance of accurately capturing the full range of observable movements to account for state- and motor-related effects on neural activity.

# Relating rich behavior to neural activity by studying ANNs

A new approach to relate neural dynamics to behavior is the use of ANNs for behavioral modeling as *artificial model organisms*. ANNs combine simple, nonlinear computational units connected together with adjustable weights, in direct analogy to the neurons and synapses in living brains [73,74]. Recurrent neural networks (RNNs) also contain recurrent feedback, again in analogy with brain anatomy. Usually, ANN weights are adjusted through an iterative learning algorithm to improve performance over many behavioral trials. If given enough examples, a ANN can perform extremely well at the trained task, often with dramatically improved performance over hand-designed solutions. After training, a researcher can analyze the ANN to glean how the behavioral task was implemented by the network.

An advantage of artificial over biological model organisms is that one may train and study thousands of networks, enabling the study of large ensembles of solutions to a given behavioral task. One powerful use of ANNs has been to generate hypotheses for how electrophysiological recordings might subserve an animal's behavior [4,14–19]. If the internal dynamics of a fully trained ANN can explain a large amount of variance in animal neural recordings, these recordings are likely to be related to the animal's task performance. Subsequently, the ANN can be studied or reverse-engineered to yield novel hypotheses about *how* neural dynamics might support the animals' behaviors. For example, ANNs have been used to discover that grid cell representations of space arise naturally as a solution to solving the problem of path integration [75,76].

As technology matures, we expect ANNs to play an increasing role in revealing how complex neural dynamics give rise to rich animal behavior. Here, increased task complexity and more detailed measures of animal behavior have the dual role of creating new ways to interpret high-dimensional neural recordings (e.g. from Neuropixels or Ca2+ imaging) and helping to better constrain ANN solutions. A related direction is to model more ancillary behavioral data as output for the ANN, e.g. not just the choice, but estimates of brain state, body movements or latent behavioral variables. For instance, instead of producing a binary decision, ANNs can produce complex motor outputs resembling behavioral motifs seen in animals [15]. It is likely that considering these kinds of behavioral details will bring the ANNs into further alignment with neural data. Finally, animals may be asked to perform an isolated behavior in the laboratory, but naturally, the animal is performing many ongoing behaviors and the brain must support all of them. It is likely that studies of ANNs trained on multiple tasks could enrich the solutions for individual tasks and bring the networks into further alignment with the neural data [77]. For example, Yang and colleagues studied how a single network implements a large [78] or huge [79] number of cognitive and memory based tasks and found functionally specific clusters for different cognitive processes, resembling cognitive specificity of neurons in prefrontal cortex.

There are some profound conceptual differences between animals and ANNs that impact behavioral modeling and will require additional research on the theory side. One large difference is that ANNs are trained only once during an optimization process and the connection weights are not subsequently modified, while animals continually update and refine their behavior. This discrepancy seems fine for understanding "instantaneous snapshots" of animal behavior but is highly problematic for understanding how animals learn or how their neural representations evolve over time [77,80]. A related consideration is that biological brains implement both the computation underlying behavior as well as the system that enables learning of novel behaviors. ANNs, however, use externally available cost functions and optimization routines, typically written as auxiliary software, which are discarded after training. The incorporation of reinforcement learning to flexibly train ANNs might be a way to overcome this limitation and allow ANNs to uncover variable task contingencies on their own [77,81,82]. Extending this approach to large numbers of ANNs will likely enable the study of differing behavioral strategies as found in behaving animals.

## Individual differences - from averages to individuality

Distinct behavioral strategies are part of a pervasive feature in many experimental and natural behaviors: *individual differences* (Table 1). Individuality refers to specific behavioral traits that differ across animals and impacts their responsiveness to the environment. For example, even animals with a similar genetic background respond differently to pharmacological interventions [83] and stress [84], and display idiosyncratic behavioral strategies during decision-making [20,50,85,86].

Recognizing the role of individual behavioral strategy can significantly change the interpretation of neural recordings and perturbations during decision-making. Mice discriminating textures show activation of different cortical areas corresponding to distinct active or passive movement strategies [85]. Consequently, cortical inactivation only affects behavior underlying the corresponding strategy. Behavioral traits also vary over an animals life time: changing the social environment reshapes activity of midbrain dopamine neurons and animal strategy in a foraging task [50].

A recent success in relating individual choice behavior of rats in an auditory discrimination task to neural activity came from RNNs [20]. Here, the internal dynamics of different random RNN instantiations matched neural dynamics from medial frontal cortex recordings, with choice selectivity emerging more strongly in high-performing rats and RNNs. Intriguingly, the stability of the RNN's internal dynamics could also be measured without sensory stimulation, providing insights beyond the recorded neural activity [20]. Future efforts will show if individual behaviors can be fit with custom ANN architectures [87], rather than using random network instantiations, to create models for individual animals' neural and cognitive dynamics.

## Conclusion

We have described how complex task-design and in-depth behavioral quantification can be leveraged to gain insights into the interplay between behavioral output and the underlying neural activity. Rather than trying to constrain behavior and focusing on a few instructed movements in simple tasks, we argue that future studies will strongly benefit from embracing diversity as a feature of animal behavior to understand previously unknown features of neural activity. More detailed behavioral information should also extend into the realm of ANNs and might provide new ways to create stronger links between artificial and living neural networks.

**Highlighted references**

Vinck et al., 2015:
Brain state changes in mice were measured through self-initiated locomotion and pupil size. Pupil-linked state transitions were present in quiescent episodes and had effects on neural activity in visual cortex that were different from locomotion. Pupil dilations coincided with suppressed firing rates and enhanced visual responses while locomotion resulted in an overall increase in firing rates.

Musall et al., 2018:
A linear regression model combined video-based movement representations and task variables to predict changes in cortex-wide neural activity during decision-making. Uninstructed movements accounted for most of the neural variance, outperforming instructed movements or task variables. Model predictions accounted for a large amount of trial-to-trial variability and could be used to identify the impact of movements on trial-averaged data.

Gilad et al., 2018:
Animal movements during a texture-discrimination task reflected an active/passive strategy, leading to remarkably different cortical activity patterns in a subsequent delay period. Correspondingly, optogenetic inactivation of cortical areas had highly variable effects on task performance that were explained by the animal's behavioral strategy.

Kurikawa et al., 2018:
Rats in an auditory detection task exhibited individual variability in their responses to unfamiliar stimuli, reflected in the stability of neural dynamics in medial frontal cortex (MFC). Multiple RNN models of MFC with randomly initiated recurrent connections recapitulated these individual behavioral and neural patterns.

Pereira et al. 2019:
This toolbox for automated animal pose tracking requires a hand-labeled training set (~100 frames) to achieve high prediction performance in flies or mice. Unsupervised dimensionality reduction of movement trajectories and subsequent clustering could identify ~20 distinct behavioral motifs of fruit fly behavior, e.g. different grooming patterns.

Yang et al. 2019:
Single RNNs were trained to perform 20 different cognitive tasks. After training, recurrent units formed clusters that were specifically tuned for different cognitive processes. The RNNs also showed mixed task selectivity, where some tasks could be solved by combining other task instructions. Training RNNs sequentially substantially increased such mixed task representation and resembled neural responses in monkey prefrontal cortex.